\documentclass[referee]{raa}           
\usepackage{graphicx,times}
\usepackage{natbib}
\usepackage{amssymb,amsmath}
\usepackage{hyperref}
\usepackage{multirow,multicol,arydshln,booktabs}

\newcommand{\rxte}{{\it RXTE}}

\newcommand{\rosat}{{\it ROSAT}}

\newcommand{\xmm}{{\it XMM}-Newton}
\newcommand{\sax}{{\it BeppoSAX}}
\newcommand{\swift}{{\it Swift}}
\newcommand{\xrt}{{\it Swift}-XRT}

\newcommand{\nustar}{\textit{NuSTAR}}
\newcommand{\hst}{{\it HST}}
\newcommand{\gaia}{{\it Gaia}}
\newcommand{\chisq}{$\chi^{2}_{\nu}$}
\newcommand{\chit}{$\chi^{2}$}

\begin{document}

   \title{Revisiting the X-ray Emission of the Asynchronous Polar V1432 Aql}
   \volnopage{ {\bf 2021} Vol.\ {\bf X} No. {\bf XX}, 000--000}
   \setcounter{page}{1}

   \author{Qishan, Wang
      \inst{1,2,3}
   \and Shengbang, Qian
      \inst{1,2,3,4}
   \and Liying, Zhu
      \inst{1,2,3,4}
   }
   \institute{
            Yunnan Observatories, Chinese Academy of Sciences (CAS), P. O. Box 110, 650216 Kunming, China; {\it wangqs@ynao.ac.cn}\\
        \and
            Key Laboratory of the Structure and Evolution of Celestial Objects, Chinese Academy of Sciences, P. O. Box 110, 650216 Kunming, China\\
        \and
            University of Chinese Academy of Sciences, Yuquan Road 19\#, Sijingshang Block, 100049 Beijing, China\\
        \and
            Center for Astronomical Mega-Science, Chinese Academy of Science, 20A Datun Road, Chaoyang Distric, Beijing, 100012, P.R. China\\
\vs \no
   {\small Received 2021 Month Day; accepted 2021 Month Day}
}

\abstract{As the only eclipsing asynchronous polar, V1432 Aql provides an excellent laboratory to study the interaction between the accreted matter and the magnetic field.  Here, we report an analysis of the X-ray data from the contemporaneous \nustar\ and \xrt\ observations. The X-ray data present a  profile with a low-intensity state for almost half an orbital period, a dip at 0.6 phase, and a peak at 0.75 phase, which suggests that there was only one accretion region during the observation and the claim is supported by the spectral analysis. The comparison with the previous data indicates that the X-ray data have an orbital modulation, as the case in \sax, rather than a spin one observed in \rosat. We attribute the orbit and spin modulations to the different accretion geometries at work. The spectral analysis of the wide-band data presents a significant reflection effect, a commonly observed soft X-ray temperature, and  the energy balance in V1432 Aql . Additionally, we obtained a low total accretion rate of 1.3 $\times$ 10$^{-10}$ M$_{\odot}$ yr$^{-1}$ and a high specific accretion rate of 3.8 g cm$^{-2}$ s$^{-1}$ which explains the strong reflection from the surface of the white dwarf. However, due to its complex emission, a more physical understanding of its accretion geometry is still outstanding.
}

   \authorrunning{Qishan, Wang et al.}             
   \titlerunning{X-ray Emission of V1432 Aql}  
   \maketitle

\section{Introduction}\label{sec:intro}
~~~~V1432 Aql (RX J1940.1-1025) is a polar (AM Herculis star) in which a highly magnetic ($\sim$ 10-200 MG) white dwarf (WD) accretes material from a low-mass, Roche lobe-filling red dwarf. Normally the magnetic torque between two component stars forces the WD to spin synchronously with its orbital rotation, but five polars are well confirmed to be slightly asynchronous, such as V1432 Aql, BY Cam, V1500 Cyg, ID Cnd and 1RXS J083842.1-282723 (hereafter J0838). For these the WD spin period $P_{spin}$ differs from the orbital period $P_{orb}$ by a few per cent, and these system are known as asynchronous polars (APs). In APs, the orientation of a WD with respect to its companion will change continuously across the beat period $P_{beat}$ ($=|1/P_{orb}-1/P_{spin}|^{-1}$) during which the material is threaded by different magnetic field lines and impacts nearest magnetic poles \citep{Wynn1992power_spectra}. This pole-switching effect confuses the identification of periods in APs.  Accretion in APs is often found to be confined to a narrow region, as in polars, e.g. in CD Ind $\text{\citep{Hakala2019MNRAS.486.2549H, Littlefield2019CDInd_TESS}}$. \cite{Wang2020ApJ...892...38W} developed a simple model in which they considered the variable intensity of accretion regions due to the departure of accretion regions from the threading point across the beat period and recognized the periods for CD Ind, BY Cam and J0838. We refer to the kind of accretion as the  stream-dominated accretion. In this accretion geometry, the emission humps are stable across the beat cycle on the orbit-folded light curve, as the case in CD Ind \citep{Hakala2019MNRAS.486.2549H, Littlefield2019CDInd_TESS}, BY Cam \citep{Silber1997MNRAS.290...25S} and J0838 \citep{Rea2017J0838-2829, Halpern2017J0838-2829}, while the spin-folded optical and X-ray light curves of V1432 Aql often exhibit two or three persistent humps along the beat phase \citep{Watson1995MNRAS.273..681W, Patterson1995PASP..107..307P, Rana2005ApJ...625..351R},  indicating a different accretion geometry may be at work in V1432 Aql,  where two or three accretion region are active at the same time. In fact, the threading of material by the magnetic field in APs is much more ineffective than that in normal polars . For example, the Doppler tomography and X-ray observations show an azimuthally extended accretion curtain in V1432 Aql, indicating that  maybe the accretion stream travels most of the way around the WD before it is fully threaded by the field lines \citep{Schwope2001LNP...573..127S, Mukai2003ApJ...597..479M}. According to this model, \cite{Littlefield2015MNRAS.449.3107L} explained the orbital modulations of the mid-eclipse time and magnitude. Moreover, the ineffective threading causes an unstable accretion between the two component stars and a cycle-to-cycle variation in the light curve.  As the only eclipsing AP, the orbital period of V1432 Aql is well established from the  stable eclipse profiles in the optical and X-ray light curves and from the radial velocity measurements \citep[e.g.][]{Patterson1995PASP..107..307P, Watson1995MNRAS.273..681W, Friedrich1996A&A...306..860F, Mukai2003ApJ...597..479M} The WD spin period is also seen from the out-of-eclipse waveform in the optical \citep{Patterson1995PASP..107..307P} as well as the X-ray data \citep{Friedrich1996A&A...306..860F, Geckeler1997A&A...325.1070G}. The orbital period is 12116.284 s, and the spin period is $\sim$ 3\% longer at 12150.3 s, with the beat period of $\sim$ 50 days \citep{Littlefield2015MNRAS.449.3107L}. It is worth noting that the strong magnetic torque should synchronize the WD on a short timescale, and the spin period of V1432 Aql was found to decrease at a rate of $\sim -0.5...-1\times 10^{-8}$ s/s, thus the estimation of the synchronization time-scale is $\sim 100...200$ years \citep{Geckeler1997A&A...325.1070G, Staubert2003A&A...407..987S, Andronov2006A&A...452..941A, Littlefield2015MNRAS.449.3107L}. Although the X-ray pulse profiles of V1432 Aql are more prominent in the spin-folded \rosat\ data \citep[0.1-2.4 keV; ][]{Geckeler1997A&A...325.1070G},  the \sax\ data (1.5-8.0 keV) are modulated on the orbital period  implying that X-ray modulation on the spin period maybe apply only to soft X-rays \citep{Mukai2003ApJ...597..479M}.  If the soft X-rays are  a result of reprocessing of the hard X-rays by the surface of the WD , all X-rays should modulate on the same period. In this paper, we show that the different X-ray modulations are associated with the different accretion geometries, and the conversion between them may be induced by the unstable accretion in V1432 Aql.

Other striking aspects surrounding V1432 Aql include the significantly high soft X-ray temperature $\sim$70 - 90 keV and the strong hard X-ray emission \citep{Staubert1994A&A...288..513S, Rana2005ApJ...625..351R} in comparison with that observed in most other polars. The accreted material from the secondary follows a free trajectory until it is channeled by the magnetic field, then falls onto WD leading to a shock, which converts the kinetic energy into the thermal energy, and forming an accretion column. The cooling of the post-shock plasma is dominated by the multi-temperature bremsstrahlung emission in hard X-rays, and the cyclotron radiation in optical and infrared \citep[see][]{Cropper1990Polars}. Thus, information on the accretion geometry is encoded in the intrinsic emission. However, the observed radiation in polars has a significant contribution from the X-rays that interact with the environment, including the surface of the WD, the pre-shock material, accretion stream and any circumstellar medium that might exist. The WD surface around the accretion regions is heated by the hard X-rays and emits the X-rays in soft energy, and the X-rays reflected by the stellar surface produce the 10-30 keV Compton reflection hump and the Fe K$\alpha$ fluorescent line at 6.4 keV, while the latter can also originate from intrinsic absorbers. And the emitted X-rays may be viewed through the complex absorbing material, which causes an absorption effect in additional to that of the Galactic interstellar medium \citep[ISM,][]{Mukai2017PASP..129f2001M}. Moreover, the reprocessing, or thermalization from blobby accretion in the bombardment regime, could contribute  a black-body component with $T_{\text{eff}}$ of 20 - 40 eV. Thus the exact measurements of the neutral hydrogen column density and distance to the system are necessary for the convincing determination of the soft X-ray emission, and it is also very important to use wide-band data to establish the characteristics of the accretion given that both the Compton reflection and complex absorption harden the spectrum. The neutral hydrogen column density is $\sim1.3\times10^{21}~ \text{cm}^{-2}$ estimated from a visual extinction of \hst\ UV spectrum \citep{Schmidt2001HSTSpectroscopy}, the distance to V1432 Aql of $456 \pm 7$ pc is calculated from \gaia\ EDR3 parallax \citep[$2.193 \pm 0.036$ mas;][]{Gaia2016A&A...595A...1G, Gaia2020arXiv201201533G}, and the ISM N$_H$ column density is consistent with the \gaia\ distance $\text{\citep{Fruscione1994ApJS...94..127F}}$ and the estimate (1.03$\times10^{21}$ cm $^{-2}$) of the HI4PI Map $\text{\citep{HI4PI2016A&A...594A.116H}}$, and the simultaneous \nustar\ and \xrt\ observations in a wide energy band provide an excellent opportunity to exploring the accretion geometry in V1432 Aql.

V1432 Aql is a magnetic cataclysmic variable included in the \nustar\ legacy survey, \cite{Shaw2020MNRAS.498.3457S} derived its WD mass by fitting to the 20-78 keV spectrum (FPMA + FPMB). In this paper, we present the analysis of archival contemporaneous \nustar\ and \xrt\ observations of V1432 Aql. In Section \ref{sec:obser}, we detail the reduction of the data, while Section \ref{sec:analy} presents the results from the analysis of the X-ray data. Finally, Section \ref{sec:concl} presents a discussion about the implications of the X-ray data and characteristics of the accretion in V1432 Aql.

\section{Observations and Data Reduction}\label{sec:obser}
~~~~The \nustar\ satellite observed V1432 Aql on 2018 April 05 for the accumulated good time intervals of 27.1 ks spanning a total data time base of 60 ks from 05:00(UTC) to 21:39(UTC) (ObsID 3046004002), and the corresponding beat phase interval of these observations is 0.082 - 0.093, using the beat ephemeris from \cite{Littlefield2015MNRAS.449.3107L}. The data were reduced using the \nustar\ Data Analysis Software as part of \textsc{heasoft} 6.21 and latest \textsc{CALDB} files following standard procedures. We used \texttt{nupipeline} to extract source X-ray events from a circular region of a radius 30 arcsec centered on the source for the \nustar\ modules, FPMA and FPMB; for the background, we chose a 50 arcsec radius, circular and source-free region on the same detector. We generated the spectra, light curves and the corresponding response files using \texttt{nuproducts}, and used the \texttt{barycorr} tool to correct the photon arrival times to the solar system barycenter.

A \xrt\ observation was obtained almost simultaneously with \nustar\ for 6.2 ks from 04:43(UTC) to 10:56(UTC) (ObsID 00088611001). The cleaned event file was created through \texttt{xrtpipeline} in a standard manner. We used \texttt{xselect} tool to extract the source X-ray data from a circular region of a radius 20 arcsec centered on the source, and the background was obtained from a nearby source-free region of a radius 40 arcsec. For the light curves, the presence of bad columns on the detector is corrected through the use of \texttt{xrtlccorr} tool and the event arrival times are barycentric corrected. We built the ancillary matrix using the tool \texttt{xrtmkarf} and used the response matrix provided by the \xrt\ calibration team. We grouped all spectra using \texttt{grappha} to have at least 25 counts per bin. Spectral fits were done with Xspec v12.10c \citep{Arnaud1996ASPC..101...17A} by using the $\chi^2$ statistic. We simultaneously fitted to the \nustar/FPMA, \nustar/FPMB, and \swift-XRT data, and allowed the cross-normalization factor relative to \nustar/FPMA to vary to allow for the cross-calibration between the instruments. 

\section{Analysis and Results}\label{sec:analy}
\subsection{Light Curves}\label{sub:xray_lc}
\begin{figure}
\begin{center}
\includegraphics[width=\columnwidth]{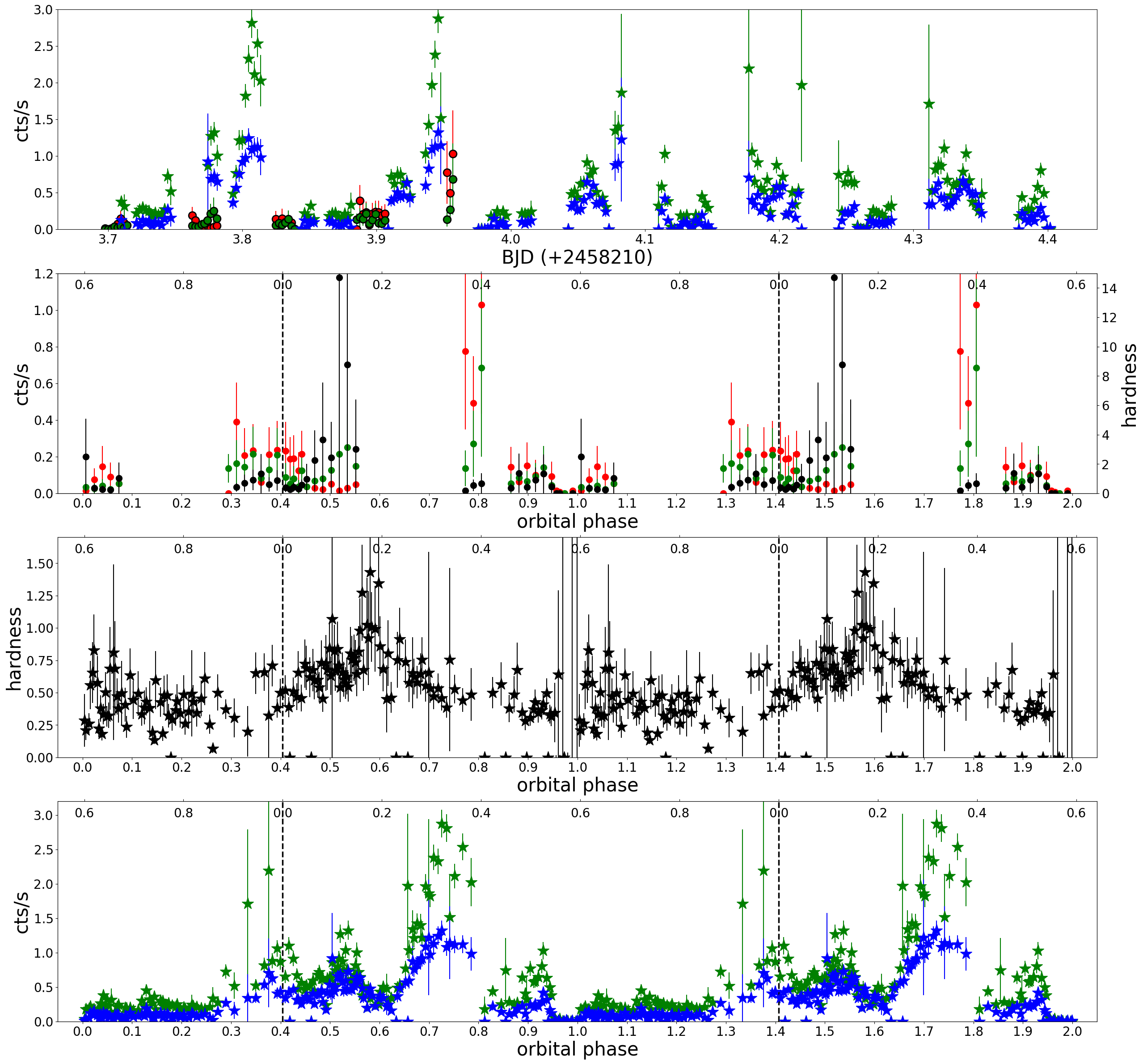}
\caption{Background-subtracted light curves in three energy bands and their hardness ratio curves. 
From bottom to top: orbit-folded \nustar\ medium (NM = 3-10 keV, the green asterisks) and hard (NH = 10-78 keV, the blue asterisks) X-ray data; their hardness ratio curve (HR1 = NH/NM, the black asterisks); orbit-folded \xrt\ soft (SS = 0.3-3 keV, the red dots), medium (SM = 3-10 keV, the green dots) X-ray data, and their hardness ratio curve (HR2 = SM/SS, the black dots); time series data for \xrt\ and \nustar.  The data are denoted, for \nustar~ (asterisks) and \swift~ (circles), with same colour scheme. We also indicate the spin phase as a twin x-axis in the lower three panels, with the spin minimum marked by vertical dashed lines. The bin size is 200s and the orbital phase is repeated for clarity. 
} \label{fig:lc}
\end{center}
\end{figure}

~~~~Fig. \ref{fig:lc}  shows the background-subtracted X-ray data in three energy bands for \xrt\ and \nustar. The energy bands - 0.3-3 keV (soft), 3-10 keV (medium) and 10-78 keV (hard) - were chosen for easy comparison between the two telescopes. The 3-10 keV data for \xrt\ have lower count rates than the corresponding \nustar\ data. However, their time series data (the top panel) indicate that the medium X-ray data are consistent within the error bars.

Although the duration of the X-ray data in the beat phase interval of 0.082-0.093 is much shorter than the 50 days beat period, the light curves are interesting in their own right. In the bottom  two panels, we folded the \nustar\ medium (NM; the green asterisks), hard (NH; the blue asterisks) X-ray data and their hardness ratio (HR1 = NH/NM; the black asterisks) on the orbital ephemeris taken from \cite{Littlefield2015MNRAS.449.3107L}. The orbital fold can be considered a spin fold with an offset ($\sim 0.4$), and we indicated the location of the spin phase 0.0 during this observation according to the optical spin ephemeris also taken from \cite{Littlefield2015MNRAS.449.3107L}. Their initial epochs were converted from HJD (UTC) to BJD (TBD) with the help of the online tool\footnote{\url{http://astroutils.astronomy.ohio-state.edu/time/hjd2bjd.html}} within the error of 1s \citep{Eastman2010PASP..122..935E}. The folded light curves have a strong peak around the orbital phase $\phi_{orb} \sim$ 0.75 and a prominent dip structure during $\phi_{orb} \sim$ 0.5-0.7 which shows up as an excess in  the hardness ratio. In addition, a broad low-intensity state during $\phi_{orb} \sim$ 0.8-0.3 (i.e. 0.8-1.0 and 0.0-0.3) for half an orbital period can be seen on the light curves, and the corresponding hardness ratio is nearly stable at a value of $\sim$ 0.4. If these features are the spin ones, a dip at the spin phase $\sim$ 0.2 and a peak at spin phase $\sim$ 0.35 are in contrast to the previous  \rosat\ and \xmm\ observations that  peak at  spin phase $\sim$ 0.2 and  dip at  spin phase $\sim$ 0.35 \citep[see the Fig. 4 in][]{Rana2005ApJ...625..351R}. The \sax\ data show two energy-dependent dip transitions which manifest themselves as two periods of enhanced hardness ratio at $\phi_{orb} \sim$ 0.2 and 0.62 \citep[see the Fig. 7 in][]{Mukai2003ApJ...597..479M}.  The dip during $\phi_{orb} \sim$ 0.5-0.7 on the \nustar\ data is corresponding to the dip at $\sim$ 0.62 on the \sax\ data, and these features on the \nustar\ data should be orbital phenomenon and hence should be  persistent across the beat cycle, as indicated by the Fig.8 of $\text{\cite{Mukai2003ApJ...597..479M}}$ and the Fig.5 of $\text{\cite{Wang2020ApJ...892...38W}}$. The \sax\ data presented two periods of enhanced hardness ratio in an orbital period, while the \nustar\ data in an orbital period presented only one period of enhanced hardness ratio. The comparison between these two X-ray data sets  indicates that only one accretion region was active during the \nustar\ observation, which is reminiscent of the case in CD Ind, BY Cam and J0838, and hence indicates that  a similar accretion geometry may  also work in V1432 Aql.  To explain the spin modulation, we adopt the picture proposed by \cite{Mukai2003ApJ...597..479M} in which V1432 Aql may have a ring around the WD. The picture is supported by the Doppler tomography \citep{Schwope2001LNP...573..127S} and perhaps explains the optical modulations of the mid-eclipse time and magnitude \citep{Littlefield2015MNRAS.449.3107L}. Systems with a  ring-dominated accretion should present a spin modulation of the light curve \citep{Wynn1992power_spectra}. Therefore, we speculate that the orbit and spin modulations on the X-ray are caused by  stream-dominated and ring-dominated accretion, respectively, and the switch between the two accretion geometries may be induced by the unstable accretion in V1432 Aql. A detailed model of the light curves for the two accretion geometries is beyond the scope of this paper, and additional  extended observations which  sample the beat cycle are  needed for such an analysis. 

The  second panel of Fig. \ref{fig:lc} shows the \xrt\ data in soft (SS = 0.3-3 keV; the red dots), medium (SM = 3-10 keV; the green dots) energy bands and their hardness ratio (HR2 = SM/SS; the black dots). The less coverage on the orbital period prevents a further analysis, and the X-ray data were used only for the spectral analysis. The X-ray eclipse, seen during $\phi_{orb} \sim$ 0.95-1.0 in the lower two panels, presents a obvious phase offset that is so significant in comparison with the amplitude of the variation of the eclipsing time \citep[$\sim$ 300 s;][]{Littlefield2015MNRAS.449.3107L}.

\subsection{Spectral Model}\label{sub:spec_model}
~~~~~~In this section, we analyzed the spectra based on three families of models (A, B, and C). A is combination of black body and plasma emission (modeled either by CEMEKL or MKCFLOW) with two components of absorption (constant ISM and local partially covering), B is the same as A, with reflection included; C is the same as B, with addition of Fe K emission.

We fitted the \xrt\ (0.3-10 keV) and \nustar\ (FPMA + FPMB; 3-78 keV) spectra simultaneously, but first excluded the Fe K region (5.5-7.5 keV). The expected multi-temperature shock structure $\text{\citep{Done1998MNRAS.298..737D}}$ indicates that the X-ray radiation can be approximated by the multi-temperature thermal emission model \textsc{cemekl} \citep{Singh1996ApJ...456..766S} based on a thermal plasma code \textsc{apec} ($switch = 2$). We chose that each temperature component is weighted by its cooling time \citep[$\alpha =1$;][]{Aizu1973PThPh..49.1184A}. The $aspl$ abundance table \citep{Asplund2009ARA&A..47..481A} was applied with all elements heavier than $\text{He}$ scaled together. The redshift was assumed to be $1.064\times10^{-7}$ from the latest estimate of the distance to V1432 Aql. The number density of hydrogen, which has little effect to the fit, was fixed to the maximum value allowed by the model ($n_{H}=10~\text{cm}^{-3}$). Since polars are strong soft X-ray emitters, we also added a blackbody component to model the low energy data.

The interaction of the radiation with matter was accounted for by the following models. In addition to a single absorber (\textsc{phabs})  for ISM, we used a usual partial-covering absorber model (\textsc{pcfabs}) since the radiation may be viewed through complex absorbers (model $A_C$: \textsc{constant$\times$phabs$\times$pcfabs$\times$(bbody+cemekl)}). The recent studies of magnetic CVs using \nustar\ have yielded very 
convincing results using the \textsc{mkcflow} model for the shock column $\text{\citep[e.g.][]{Lopes2019ApJ...880..128L}}$. For comparison, we also employed \textsc{mkcflow} for some fits to model the same data (i.e. model $A_M$: \textsc{constant$\times$phabs$\times$pcfabs$\times$(bbody+mkcflow)}).

When investigating the reflection of the WD surface, we applied the \textsc{reflect} model \citep{Magdziarz1995MNRAS.273..837M} with the same abundances as the X-ray emitting plasma assuming freshly accreted matter covers its surface (model  $B_C$: \textsc{constant$\times$phabs$\times$pcfabs$\times$(bbody+reflect$\times$cemekl)} and $B_M$: \textsc{constant$\times$phabs$\times$pcfabs$\times$(bbody+reflect$\times$mkcflow)}). We fixed the unconstrained inclination angle of the reflecting surface at the default value of \textsc{reflect} with $\cos{\mu}~=~0.45$ as having a spectral shape close to the ensemble average $\text{\citep{Magdziarz1995MNRAS.273..837M}}$. In order to test the significance of the reflection effect, we followed $\text{\cite{Mukai2015ApJ...807L..30M}}$ to fit the data once with the reflection model excluded (model $A_C$ and $A_M$), and once with it included (model $B_C$ and $B_M$).  For a wider investigation of model parameter space, the ISM density was let free during these fits.

In Table \ref{tab:spectral_results_all}, we report the best-fit results with their 90\% confidence error ranges and the goodness of fit  for the various models considered, and the difference in $\chi^2$ ($\Delta\chi^2$) compared to the no-reflection case in the last  column. The large $\Delta\chi^2 > 60$ and the unphysically high reflection amplitude $>$ 3 indicate that the reflection effect is highly significant, hence imply a low shock height and a high specific mass accretion rate in V1432 Aql \citep{Mukai2015ApJ...807L..30M}. Next we modeled all data in the 0.3-78 keV range, and introduced a Gaussian emission  line to represent the fluorescent Fe K$\alpha$ line at 6.4 keV (model  $C_C$: \textsc{constant$\times$phabs$\times$pcfabs$\times$(bbody+gauss+reflect$\times$cemekl)} and $C_M$: \textsc{constant$\times$phabs$\times$pcfabs$\times$(bbody+gauss+reflect$\times$mkcflow)}). The low accretion column means that at most half of radiation of the post-shock plasma should be intercepted and reprocessed by the surface, so we fixed the reflection amplitude to 1 ($rel_{refl}=1$, the maximum value with physical meaning) while allowed $\cos{\mu}$ to vary, also we fixed the abundance to the value from the previous best fit because the Gaussian line could severely affect it. Except for the  fits of these models with free ISM density, we also fitted the models with the ISM density fixed at $1.3\times10^{21}~\text{cm}^{-2}$ estimated from a visual extinction of \hst\ UV spectrum $\text{\citep{Schmidt2001HSTSpectroscopy}}$. We summarized the best-fit results in  Table \ref{tab:spectral_results_all}, and shown the spectra for the trial $C_C\#2$ in Figure \ref{fig:spectral_fit}.  The  fits based on the same model present a consistent value for all parameters. Therefore, for  convenience, we focused only the results from the trial $C_C\#2$ in the following analysis. The low orbit-averaged $\mu$ ($<37^{\circ}$) suggests a face-on viewing geometry, which implies that the latitude $\beta$ of the accretion regions is roughly equal to the inclination $i$ of the system. Also the \nustar\ spectra suggest a pronounced fluorescent iron line at 6.4 keV with an equivalent width (EW) estimated at $201_{-52}^{+61}~\text{eV}$. The reflection can contribute EW as much as $\sim 100 ~ \text{eV}$, and the partial absorber with $N_{\text{H,eff}} \sim 9 \times 10^{22} ~\text{cm}^{-2}$, defined as the sum of $C_F\times\text{N}_H$ for each partial absorber, could account for the rest $\sim$ 100 eV \citep[$=N_{\text{H,eff}}/(10^{21} ~\text{cm}^{-2})$ eV;][]{Ezuka1999ApJS..120..277E}. We conclude that the shock-front has to form immediately above the WD surface in order to explain the significant reflection and the strong Fe K$\alpha$ 6.4 keV line (see Sec.~\ref{sub:compa}).

\begin{table}
\centering
\renewcommand\tabcolsep{3pt}
\caption{Best-fit Results for the Averaged Spectra.} \label{tab:spectral_results_all} 
\begin{tabular}{lcccccccccccc}
\hline\hline
Trial & \textsc{phabs} & 
\multicolumn{2}{c}{\textsc{pcfabs}} & \multicolumn{1}{c}{\textsc{bbody}} & \multicolumn{1}{c}{\textsc{gauss}} &
\multicolumn{2}{c}{\textsc{reflect}} &
\multicolumn{2}{c}{\textsc{cemekl/mkcflow}} &
\multicolumn{1}{c}{\chit/d.o.f. (\chisq)} &
\multicolumn{1}{c}{$\Delta\chi^2$$^\dagger$}\\ 
\cmidrule(r){3-4} \cmidrule(r){7-8} \cmidrule(r){9-10} & $N_{H1}$ &  $N_{H2}$ & $C_F$ & $kT$ & $\sigma$ & $rel_{refl}$ & $\cos{\mu}$ & $kT_{max}$ & $Z$ & &\\
\cmidrule(r){2-3} & \multicolumn{2}{c}{$10^{22}~\text{cm}^{-2}$} & & eV & keV & & & keV & $Z_\odot$ \\
\hline
\multicolumn{13}{l}{$A_C$: \textsc{constant$\times$phabs$\times$pcfabs$\times$(bbody+cemekl)}}\\ 
\hline
$A_C$ & 0.29$_{-0.06}^{+0.13}$&28$_{-5}^{+4}$&0.75$_{-0.04}^{+0.02}$&38$_{-3}^{+3}$&-&-&-&42$_{-4}^{+4}$&0$_{}^{+0.43}$&502/440(1.14)&-
\\
\hline
\multicolumn{13}{l}{$A_M$: \textsc{constant$\times$phabs$\times$pcfabs$\times$(bbody+mkcflow)}}\\ 
\hline
$A_M$ & 0.34$_{-0.07}^{+0.16}$&31$_{-6}^{+3}$&0.78$_{-0.02}^{+0.02}$&38$_{-1}^{+11}$&-&-&-&47$_{-5}^{+7}$&0$_{}^{+0.29}$&515/440(1.17)&-
\\
\hline
\multicolumn{13}{l}{$B_C$: \textsc{constant$\times$phabs$\times$pcfabs$\times$(bbody+reflect$\times$cemekl)}}\\ 
\hline
$B_C$ & 0.10$_{-0.10}^{+0.12}$&8.3$_{-2.0}^{+0.1}$&0.66$_{-0.05}^{+0.06}$&53$_{-2}^{+5}$&-&3.2$_{-0.2}^{+0.2}$&0.45$^*$&34$_{-1}^{+3}$&0.22$_{-0.02}^{+0.02}$&435/439(0.99)&67
\\
\hline
\multicolumn{13}{l}{$B_M$: \textsc{constant$\times$phabs$\times$pcfabs$\times$(bbody+reflect$\times$mkcflow)}}\\ 
\hline
$B_M$ & 0.11$_{-0.11}^{+0.13}$&7$_{-1}^{+8}$&0.64$_{-0.02}^{+0.02}$&52$_{-1}^{+11}$&-&4.9$_{-2.0}^{+0.6}$&0.45$^*$&37$_{-5}^{+7}$&0.22$_{-0.07}^{+0.13}$&437/439(0.99)&78
\\
\hline
\multicolumn{12}{l}{$C_C$: \textsc{constant$\times$phabs$\times$pcfabs$\times$(bbody+gauss+reflect$\times$cemekl)}}
\\\hline
$C_C$\#1 & 0.14$_{-0.10}^{+0.11}$&12.1$_{-0.4}^{+2.1}$&0.71$_{-0.06}^{+0.03}$&49$_{-9}^{+8}$&0.16$_{-0.05}^{+0.08}$&1$^*$&0.95$_{-0.25}^{}$&37.6$_{-1.9}^{+2.0}$&0.22$^*$&554/543(1.02)&-
\\
$C_C$\#2 & 0.13$^*$&12.1$_{-2.3}^{+2.2}$&0.71$_{-0.04}^{+0.04}$&50$_{-7}^{+6}$&0.15$_{-0.09}^{+0.08}$&1$^*$&0.95$_{-0.15}^{}$&37.3$_{-2.7}^{+3.0}$&0.22$^*$&555/544(1.02)&-
\\
\hline
\multicolumn{12}{l}{$C_M$: \textsc{constant$\times$phabs$\times$pcfabs$\times$(bbody+gauss+reflect$\times$mkcflow)}}\\\hline
$C_M$\#1 & 0.21$_{-0.13}^{+0.11}$&14.1$_{-0.9}^{+1.4}$&0.73$_{-0.02}^{+0.01}$&40$_{-4}^{+2}$&0.13$_{-0.04}^{+0.08}$&1$^*$&0.95$_{-0.24}^{}$&45.3$_{-1.5}^{+2.0}$&0.22$^*$&563/543(1.04)&-
\\
$C_M$\#2 & 0.13$^*$&13.0$_{-0.5}^{+1.6}$&0.75$_{-0.02}^{+0.01}$&48$_{-10}^{+3}$&0.13$_{-0.02}^{+0.07}$&1$^*$&0.95$_{-0.26}^{}$&45.8$_{-3.4}^{+3.7}$&0.22$^*$&563/544(1.03)&-
\\
\hline\hline
\end{tabular}
\begin{minipage}{\textwidth}
\vspace{5pt}
Notes:\\
$^*$~fixed parameter, and see text for the others. \\
$^\dagger$~difference in $\chi^2$ between the fits excluding reflection and including reflection.
\end{minipage}
\end{table}

\begin{table}
\centering
\renewcommand\tabcolsep{3pt}
\caption{X-ray Flux and Luminosity of the components for different trials.} \label{tab:spectral_results_flux}
\begin{tabular}{cccccccc}
\hline\hline
Trial & 
\multicolumn{2}{c}{\textsc{bbody}} & \multicolumn{2}{c}{\textsc{reflect}} & \multicolumn{2}{c}{\textsc{cemekl/mkcflow}} &
Ratio$^\ddagger$ \\
\cmidrule(r){2-3} \cmidrule(r){4-5} \cmidrule(r){6-7}& F$^*$ & L$^\dagger$ & F$^*$ & L$^\dagger$ & F$^*$ & L$^\dagger$ & \\\hline
$C_C$\#1 & 
9.3$_{-8.0}^{+103}$ & 5.8$_{-5.0}^{+66}$ & 
2.2$_{-0.4}^{+0.3}$ & 1.4$_{-0.3}^{+0.2}$ & 
4.9$_{-0.1}^{+0.2}$ & 6.1$_{-0.3}^{+0.5}$ & 
46$_{-39}^{+40}$\%\\
$C_C$\#2 & 
8.2$_{-4.2}^{+23}$ & 5.1$_{-3.3}^{+15}$ & 
2.2$_{-0.3}^{+0.3}$ & 1.4$_{-0.2}^{+0.2}$ & 
5.0$_{-0.1}^{+0.2}$ & 6.2$_{-0.3}^{+0.5}$ & 
49$_{-28}^{+27}$\%\\
$C_M$\#1 & 
25$_{-21}^{+487}$ & 16$_{-13}^{+313}$ & 
2.3$_{-0.4}^{+0.3}$ & 1.4$_{-0.3}^{+0.3}$ & 
5.3$_{-0.1}^{+0.2}$ & 6.6$_{-0.4}^{+0.4}$ & 
28$_{-26}^{+45}$\%\\
$C_M$\#2 & 
10.8$_{-7.4}^{+37}$ & 6.7$_{-4.7}^{+24}$ & 
2.3$_{-0.2}^{+0.2}$ & 1.4$_{-0.3}^{+0.2}$ & 
5.3$_{-0.1}^{+0.2}$ &  6.6$_{-0.4}^{+0.4}$ & 
45$_{-29}^{+29}$\%\\
\hline\hline
\end{tabular}
\begin{minipage}{\textwidth}
\vspace{5pt}
Notes: \\
$^*$ the  component flux (0.01 eV - 100 keV) in units of $10^{-11}$ erg s$^{-1}$ cm$^{-2}$.\\
$^\dagger$ the  component luminosity (0.01 eV - 100 keV) in units of $10^{32}$ erg s$^{-1}$.\\
$^\ddagger$ ratio of the plasma luminosity to the total luminosity, see the text for more.
\end{minipage}
\end{table}

\begin{table*}
\centering
\renewcommand\tabcolsep{3pt}
\caption{Best-fit Results of the Model  $C_C^*$ Fitted to the Orbit Phase-resolved Spectra} \label{tab:spectral_results_phased}
\begin{tabular}{cccccccc}
\hline\hline
Phase       & \multicolumn{2}{c}{\textsc{pcfabs}} & \textsc{bbody} & \textsc{gauss} & \textsc{reflect}& \textsc{cemekl} & \chit/d.o.f. (\chisq)\\
\cmidrule(r){2-3} & $N_{H2}$ & $C_F$ & $kT$& $\sigma$ & $\cos{\mu}$ & $kT_{max}$ & \\
            & $10^{22}~\text{cm}^{-2}$ & & eV & keV & & keV & \\
\hline
0.80-0.30 & $6.49_{-0.52}^{+0.52}$  & $0.56_{-0.03}^{+0.03}$ & $10_{-2}^{+3}$  & $0.21_{-0.01}^{+0.01}$  & $0.32_{-0.01}^{+0.01}$  & $29.15_{-0.51}^{+0.51}$  & 131.6/140(0.94)\\
0.30-0.65 & $22.95_{-2.08}^{+3.22}$ & $0.80_{-0.03}^{+0.02}$ & $44_{-8}^{+4}$  & $0.14_{-0.06}^{+0.09}$ & $0.95_{-0.25}$          & $40.62_{-2.76}^{+5.24}$  & 305.28/288(1.06)\\
0.65-0.80 & $6.46_{-0.39}^{+3.46}$  & $0.95_{-0.27}^{+0.04}$ & -             & $0.10_{-0.08}^{+0.31}$ & $0.93_{-0.18}^{+0.02}$  & $37.33_{-4.60}^{+1.19}$  & 175.77/189(0.93)\\
\hline\hline
\end{tabular}
\begin{minipage}{\textwidth}
\vspace{5pt}
Notes:\\
$^*$ the fixed parameters are the same as that of the  trial $C_C$\#2 fitted to the averaged spectra.\\
\end{minipage}
\end{table*}

\subsection{Orbit Phase-Resolved X-ray Spectra}\label{sub:orbit_spect}
\begin{figure}[htbp]
\centering
\subfloat[Spectra during $\phi_{orb}=0.0-1.0$.]
{
    \begin{minipage}[t]{0.5\textwidth}
        \centering 
        \includegraphics[width=1\textwidth]{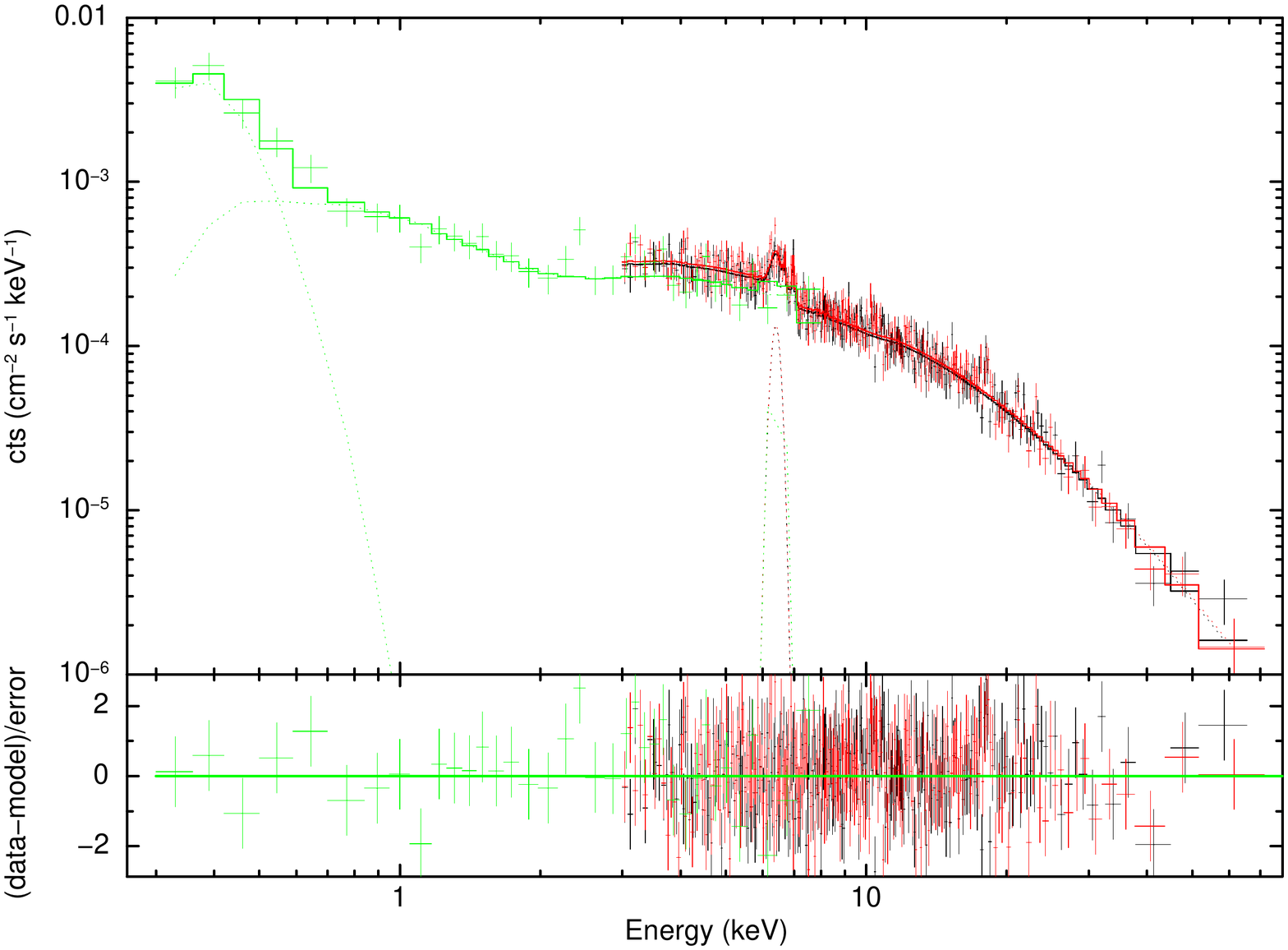}
    \end{minipage}
}
\subfloat[Spectra during $\phi_{orb}=0.8-0.3$.]
{
    \begin{minipage}[t]{0.5\textwidth}
        \centering
        \includegraphics[width=1\textwidth]{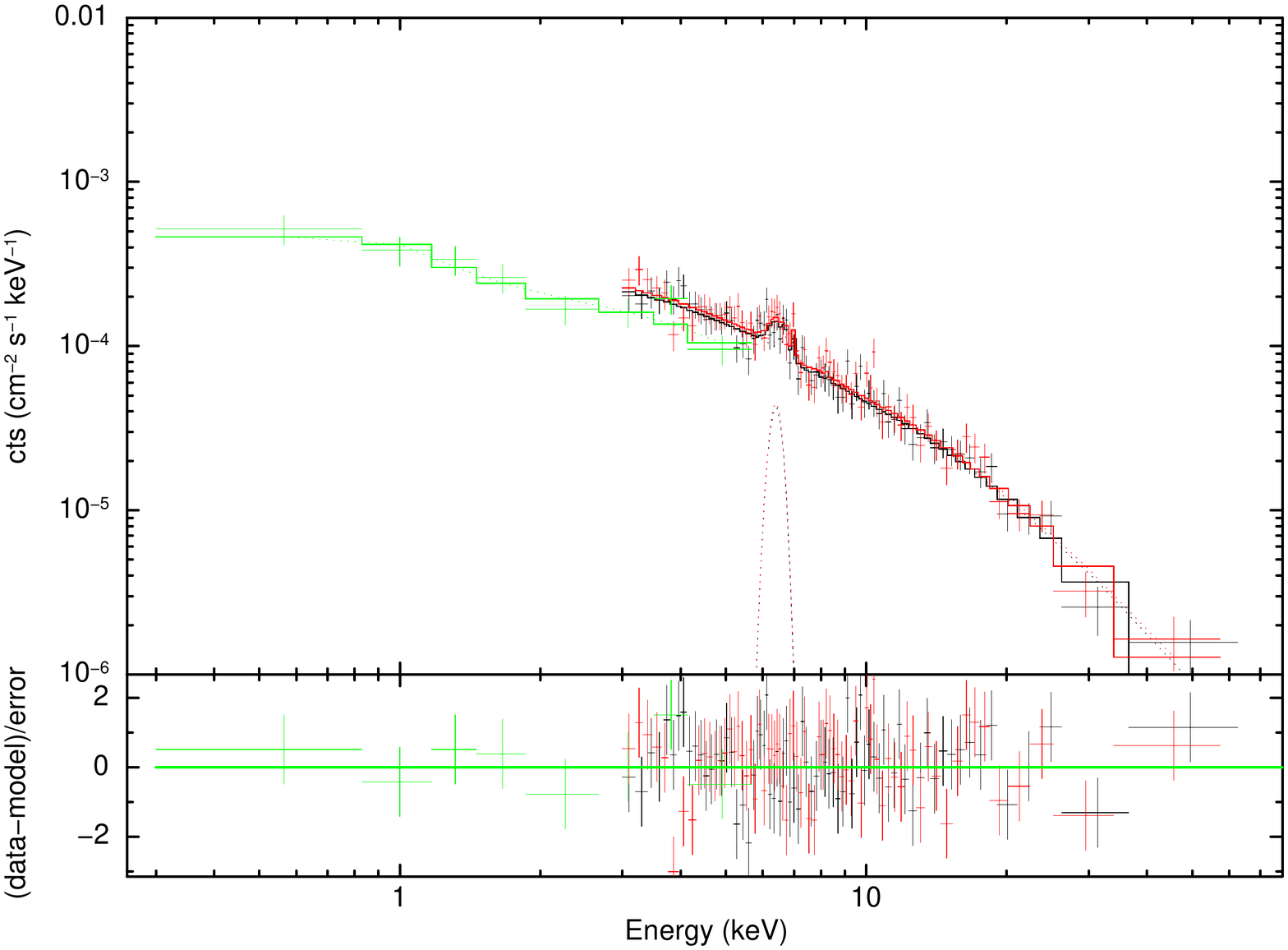}
    \end{minipage}
}

\subfloat[Spectra during $\phi_{orb}=0.3-0.65$.]
{
    \begin{minipage}[t]{0.5\textwidth}
        \centering
        \includegraphics[width=1\textwidth]{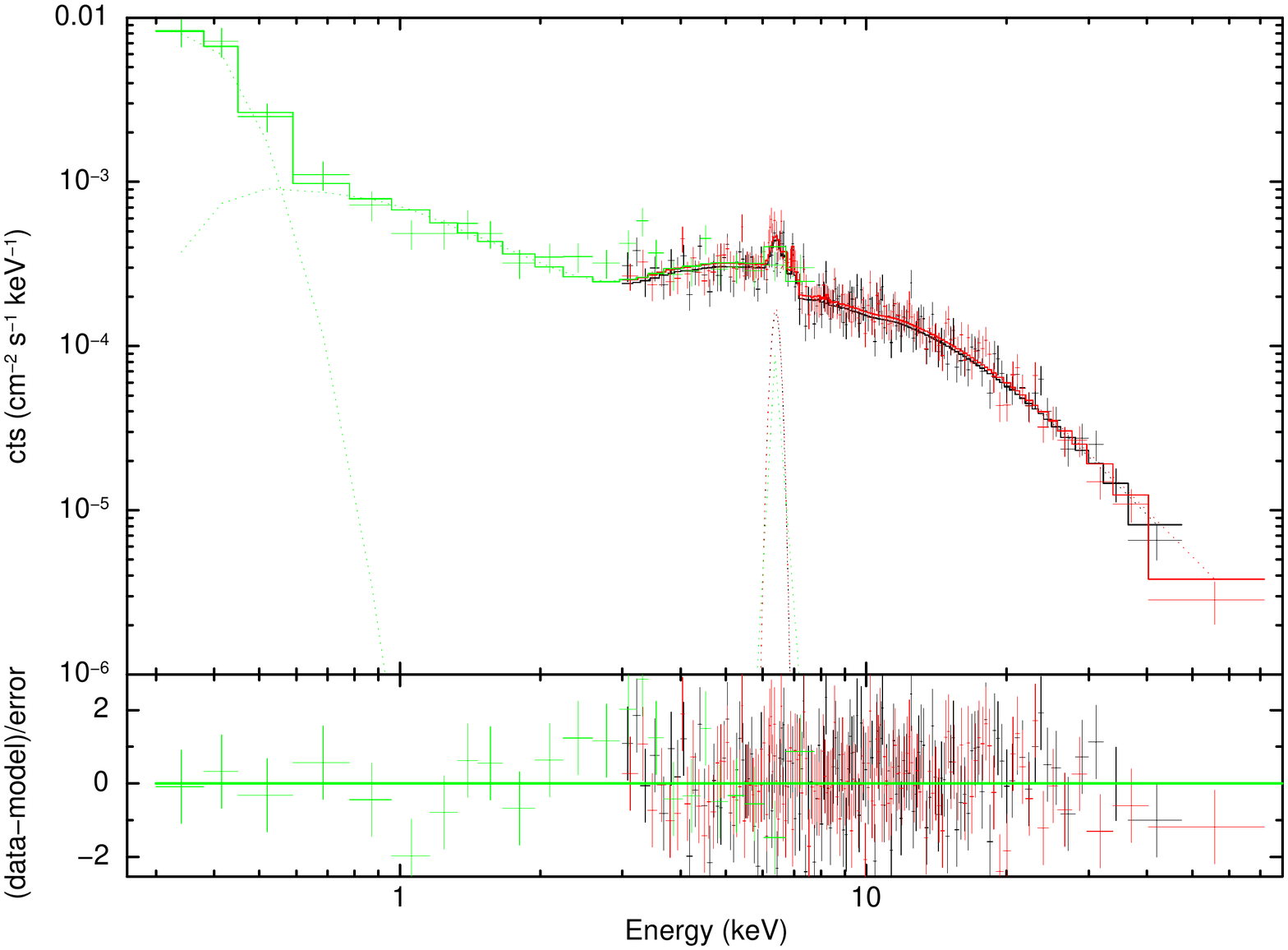}
    \end{minipage}
}
\subfloat[Spectra during $\phi_{orb}=0.65-0.8$.]
{
    \begin{minipage}[t]{0.5\textwidth}
        \centering
        \includegraphics[width=1\textwidth]{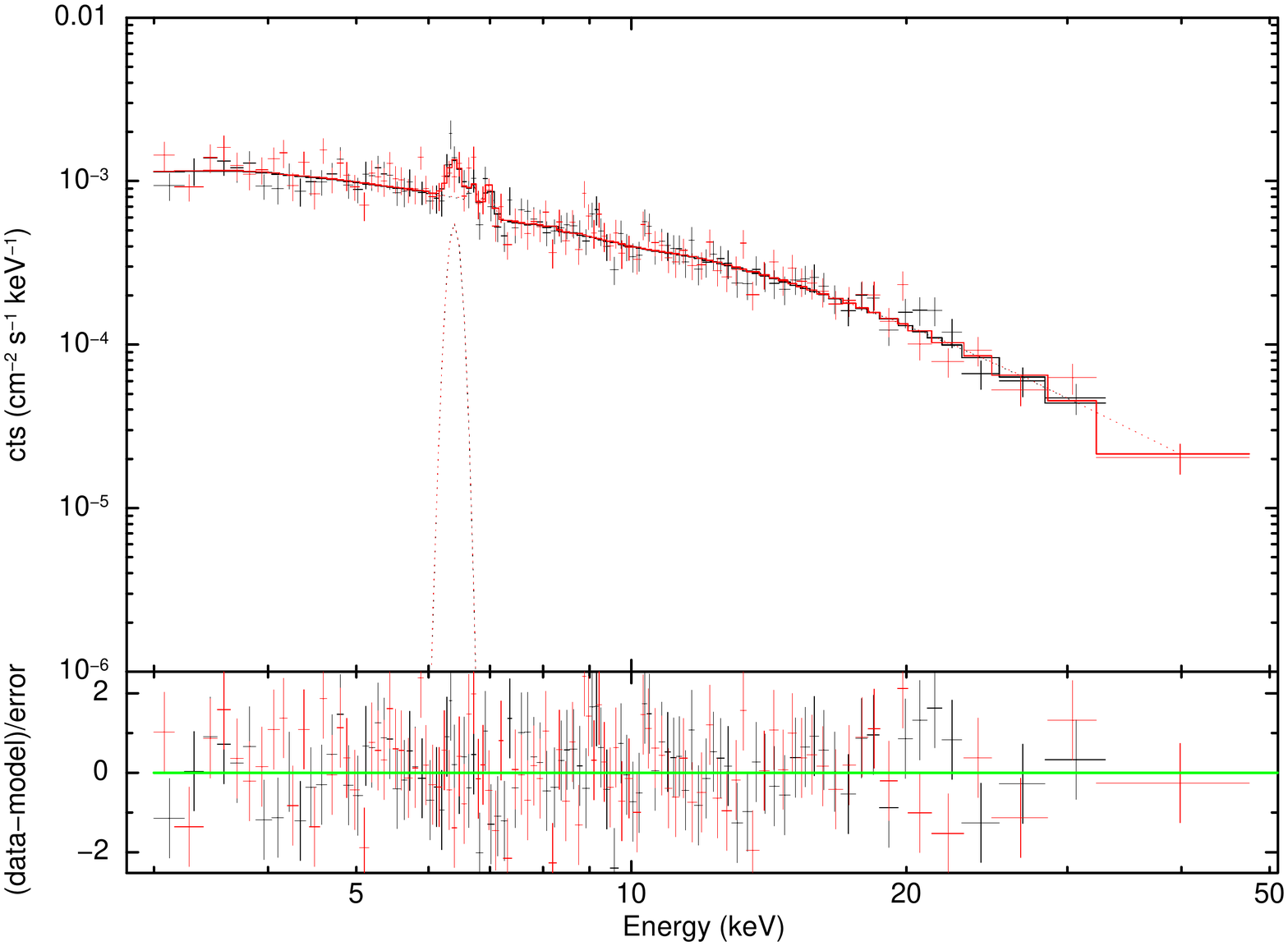}
    \end{minipage}
}
\caption{\nustar\ (black and red) + \xrt\ (light blue) spectra during the different phase intervals and their best-fit models. The spectra were modeled with a multi-temperature plasma model, plus a soft blackbody emission and a fluorescent Fe K$\alpha$ line (the model  $C_C$ with fixed $N_{H1}$), except for the data during $\phi_{orb}=0.65-0.8$ for which the blackbody component was not included in the model  $C_C$. The fit residuals in units of $\chi^{2}$ for each spectra is plotted in the bottom panels. \label{fig:spectral_fit}}
\end{figure}

~~~~To gain a better insight into the X-ray emission and the physical characteristics of the accretion regions, we analysed the orbit phase-resolved X-ray spectra during $\phi_{orb} = 0.8-0.3$, $\phi_{orb} = 0.3-0.65$, and $\phi_{orb} = 0.65-0.8$. The division is based on the analysis of the X-ray light curves and ensures the each phase interval has enough counts. The main  characteristics for each of the phase intervals are the broad emission minimum during $\phi_{orb} = 0.8-0.3$, the dip structure during $\phi_{orb} = 0.3-0.65$, and the emission pulse during $\phi_{orb} = 0.65-0.8$ (see Sec. \ref{sub:xray_lc}). The phase-resolved spectra were fitted using the model  $C_C$, except for the data during $\phi_{orb} = 0.65-0.8$ which was fitted by the model excluded the blackbody component because the \xrt\ data has little counts during the phase interval and was not used. The fixed parameters and their values are the same as in the spectral fit $C_C$\#2  (Sec.\ref{sub:spec_model}). We presented the best-fit results for the phase-resolved spectra in  Table \ref{tab:spectral_results_phased}, and plotted the fitted spectra, the best-fit model, along with the individual model component, in Fig. \ref{fig:spectral_fit}.  During the minimum phase  ($\phi_{orb}=0.8-0.3$) the accretion region is not seen, and the surface in view was not  accreting but probably was active  half a beat period before, due to pole switching, so cooled to a much low temperature ($\sim 10$ eV), while during $\phi_{orb} = 0.3-0.65$ the accretion column came into view then was aligned with the line of sight at $\sim 0.6$, which gave a face-on viewing geometry ($\cos{\mu} \sim 0.95$) and the highest absorbing column density of $N_{H2} \sim 23\times 10^{22}$ cm $^{-2}$.  This suggests that the absorption was caused by the immediate pre-shock flow \citep[see figure 8 in][]{Mukai2017PASP..129f2001M}. When entering the next phase  interval, the photons from the accretion column are less absorbed and  produce an emission peak  at $\phi_{orb} = 0.65-0.8$. The same inclination angles of the reflecting surface $\mu$ and same maximum temperature $kT_{max}$ within the error bars for $\phi_{orb} = 0.3-0.65$ and $\phi_{orb} = 0.65-0.8$ indicate that the accretion region stretches for almost half an orbital period on the WD, as inferred in the previous studies $\text{\citep[e.g.][]{Rana2005ApJ...625..351R}}$.

\subsection{The Characteristics of the X-Ray Radiation}\label{sub:xray_chara}
\begin{figure}[htbp]
\centering
\subfloat[]
{
    \begin{minipage}[t]{0.5\textwidth}
        \centering 
        \includegraphics[width=1.2\textwidth]{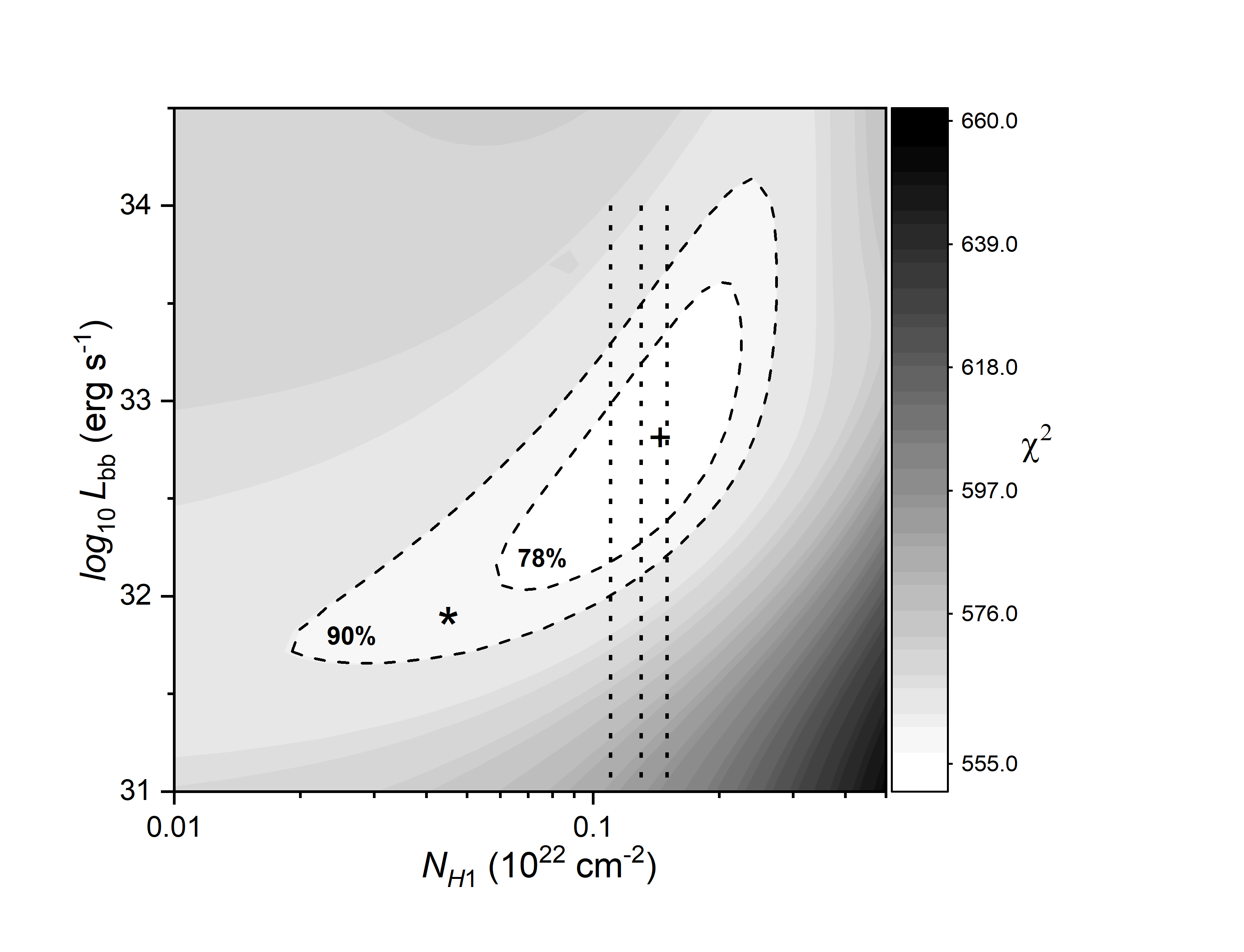}
    \end{minipage}
}
\subfloat[]
{
    \begin{minipage}[t]{0.5\textwidth}
        \centering
        \includegraphics[width=1.2\textwidth]{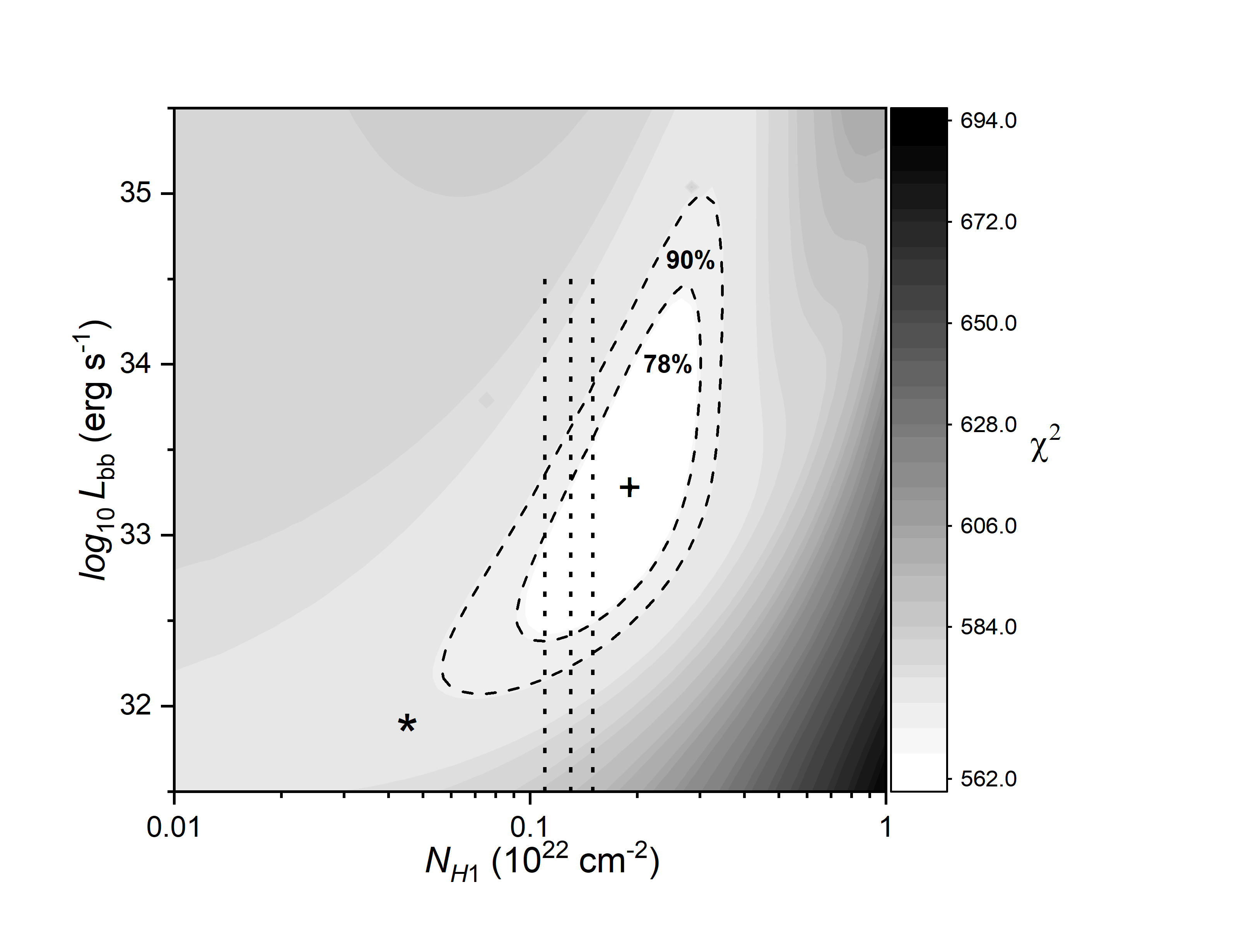}
    \end{minipage}
}
\caption{The 78 and 90 per cent confidence regions (with $\chi^2 <$ $\chi^2_{\text{min}}$+2.3 and $\chi^2_{\text{min}}$+4.6 for two interesting parameters, where $\chi^2_{\text{min}}$ is the corresponding goodness of fit taken from Table \ref{tab:spectral_results_all}) for the hydrogen column density of ISM and total blackbody luminosity allowed by the  fits $C_C\#1$ (a) and $C_M\#1$ (b). The crosses mark their best-fit positions, the asterisks indicate the $\text{\cite{Rana2005ApJ...625..351R}}$ value, and the vertical dotted lines (from the left to right in each panel) are corresponding to $N_{H1} / (10^{22} \text{cm}^{-2}$) = 0.11, 0.13, and 0.15, respectively. \label{fig:banana_plots}}
\end{figure}

~~~~The spectral fitting results imply that a substantial fraction (up to a half) of the radiation from the post-shock plasma must be intercepted by the WD surface, where the soft and medium energy X-rays are largely absorbed and reprocessed, while the hard X-rays are reflected \citep{Mukai2017PASP..129f2001M}. Therefore, in theory, the emission directly from the accretion column should account for a half of the total emission. In order to study the energy balance, we need to calculate the total unabsorbed X-ray luminosity from each component. Considering the anisotropic emission from the accretion column, we shall restrict the analysis of the energy balance to the average spectra. We extrapolated the best-fit model up to 100 keV and down to 0.01 eV, the unabsorbed flux $F_{cemekl}$ from the accretion column is derived as $5.0_{-0.1}^{+0.2} \times 10^{-11}~ \text{erg} ~\text{s}^{-1} ~\text{cm}^{-2}$ which is converted to the total luminosity $L_{cemekl} = 2\pi d^2 F_{cemekl}$, corresponding to spatial averaged emission into the half sphere at a distance $d$, of $6.2^{+0.5}_{-0.3} \times 10^{32} ~ \text{erg} ~ \text{s}^{-1}$. The soft X-ray emission comes from the WD surface and its bolometric luminosity is given by $L_{bb} = \pi d^2 F_{bb} \sec{\theta}$, where $\theta$ is the mean viewing angle to the emission region \citep{Ramsay2004MNRAS.347..497R}.  Considering the face-on accretion region, we have $\sec{\theta}=1/\cos{\mu}\approx1$, so the total blackbody luminosity $L_{bb}$ can be estimated from its time-averaged flux $8.2_{-4.2}^{+23} \times 10^{-11}~ \text{erg} ~\text{s}^{-1} ~\text{cm}^{-2}$ as $5.1^{+15}_{-3.3} \times 10^{32} ~ \text{erg} ~ \text{s}^{-1}$. Following the same procedure above, the reflection luminosity $L_{reflect}$ is calculated as $1.4^{+0.2}_{-0.2} \times 10^{32} ~ \text{erg} ~ \text{s}^{-1}$. We estimate that the luminosity outside of 0.01 eV - 100 keV is about 1\%  of the 0.01-100 keV luminosity, so the sum of these emission components can be used to estimate the total X-ray luminosity as $L_{total}=13^{+15}_{-4} \times 10^{32} ~ \text{erg} ~ \text{s}^{-1}$, and $L_{cemekl}$ occupies $49^{+27}_{-28}\%$ share of the total radiation. The other trials reveal consistent results for the X-ray flux and luminosity of these components (see Table \ref{tab:spectral_results_flux}), and we  find that V1432 Aql is consistent with the model which includes radiative heating only. Note that the soft X-ray flux of V1432 Aql is much more significant than the previous result $\text{\citep[ $8\times10^{31}\text{erg}~ \text{s}^{-1}$, corrected to the \gaia\ distance;][]{Rana2005ApJ...625..351R}}$. In order to check this result, in Fig. \ref{fig:banana_plots}, we plotted the confidence contours in the $N_{H1}/L_{bb}$ plane for the trials $C_C\#1$ and  $C_M\#1$, and marked the positions for the best-fit and the $\text{\cite{Rana2005ApJ...625..351R}}$ results. We find that the deviation between the estimates for $L_{bb}$ from $\text{\cite{Rana2005ApJ...625..351R}}$ and this work results from the different $N_{H1}$ values used. The trials favor a higher $N_{H1}$ than that used by $\text{\cite{Rana2005ApJ...625..351R}}$, and the higher $N_{H1}$ is much more consistent with the estimate from the UV dip. Therefore, the soft X-ray flux of V1432 Aql is probably higher than previously thought. 

\subsection{The parameters of the system}\label{sub:param}
~~~~A low height of shock column implies that the highest temperature $T_s$ just below the shock is proportional to the gravitational potential of the WD \citep{Aizu1973PThPh..49.1184A} as $kT_{s} = \frac{3}{8} \mu m_p G M_{1}/R_{1}$, where $\mu \approx 0.62$ is the mean molecular weight, $k$ is Boltzmann constant, $m_p$ is the proton mass, $M_{1}$ is the WD mass, and $R_{1}$ is its radius. Therefore, by combining the measured $T_{max}$ from the trials $C_C\#2$ and $C_M\#2$ with the \cite{Nauenberg1972Analytic} mass-radius relation of WDs, we can calculate $M_{1}$= 0.85 $\pm$ 0.07 $M_{\odot}$ and $R_{1}=0.01 \pm 0.0003~R_{\odot}$. Adopting the mass–radius relation for CV donor stars \citep{McAllister2019MNRAS.486.5535M} and assuming that its volume is equal to that of its Roche lobe \citep{Eggleton1983ApJ...268..368E}, we estimate the mass of the secondary as $M_2=0.22\pm{0.02}~M_{\odot}$. The X-ray eclipsing analysis \citep{Mukai2003ApJ...597..479M} gave the eclipse width of 695 s, which implies the inclination of the system as $i$=78.5$\pm$0.5$^{\circ}$. The accretion rate $\dot{M}$ can be estimated assuming that the accretion luminosity is emitted mostly in X-rays and is given by $\dot{M} = L_{total} R_{1}/GM_{1}$. In view of the energy balance of V1432 Aql, the total accretion luminosity is estimated at  $1.3(1)$ $\times$ $10^{33} ~ \text{erg} ~ \text{s}^{-1}$ from  the hard luminosity to reduce the uncertainty. Substituting these values in equation above gives a mass accretion rate of  1.3(2) $\times$ $10^{-10}~M_{\odot}yr^{-1}$. We note that this low accretion rate  rules out a post-nova classification of V1432 Aql given the short synchronization time-scale for APs \citep{Littlefield2015MNRAS.449.3107L, Schmidt1991ApJ...371..749S} and that the cataclysmic variable with an orbital period $>$ 3 hours should not enter into a deep hibernation after a nova outburst \citep{Fang2021MNRAS.501.3046F}.

\subsection{Comparison with previous results}\label{sub:compa}
~~~~The previous X-ray data for \rxte\ and \xmm\ were studied in detail by $\text{\cite{Singh2003A&A...410..231S}}$ and \cite{Rana2005ApJ...625..351R}. The \xmm\ data presented in $\text{\cite{Singh2003A&A...410..231S}}$ show orbital modulations reminiscent of the \nustar\ ones, especially the hardness ratio (=(2-12 keV)/(0.5-2.0 keV)) curve reveals a pulse at $\phi_{orb}=0.65$ that may be corresponding to the enhanced hardness ratio at $\phi_{orb}=0.65$ of the \nustar\ data. While the orbital light curves of the \xmm\ data reveal three peaks almost uniformly located over the orbital period, and the three peaks of the \nustar\ data are concentrated in half an orbital period, which is consistent with the claim that only one accretion region was active during the NuSTAR observations.  $\text{\cite{Rana2005ApJ...625..351R}}$ derived a comparatively low soft-to-hard ratio of 0.16, while the best-fit spectral results reported here for the \nustar\ + \xrt\ data yielded  less of an imbalance, consistent with both the prediction of the standard model and the result of \cite{Ramsay2004MNRAS.347..497R}. They also derived a much high blackbody temperature of $88\pm2$ eV by adopting a low neutral hydrogen column density $\sim 4.5\times10^{20}$cm$^{-2}$. With the $N_{H1}$ of \textsc{phabs} fixed to this value, \xrt\ data gave a blackbody temperature of $72 \pm 24$ eV in agreement with their value within the errors.  While the estimate of $N_{H1}$ from the absorption dip at 2200 {\AA} is consistent with that from the HI4PI Map, and the derived blackbody temperature $T_{bb}=50$ eV is commonly observed in polars \citep[20-60 eV;][]{Ramsay2004MNRAS.347..497R}.  \cite{Beuermann2012A&A...543A..41B} found that an exponential distribution of the emitting area vs. blackbody temperature is physically more plausible and can fit the soft X-rays of AM Her much better. Using the new soft X-ray spectral model, we can estimate, from $L_{bb}$ and $T_{bb}$,  the value of the fractional heated area as $f \sim 0.034 \%$ and the mean specific mass accretion rate as $\dot{m} \sim 3.8 ~g ~cm^{-2} ~s^{-1}$, a typical rate in a high state of polars. In terms of the specific accretion rate, we can predict the shock height on the WD surface of $H_{sh} \sim 0.002 R_{1}$ \citep{Frank2002Accretion}, the low height is consistent with the strong reflection detected in V1432 Aql. Although the total accretion rate of $\text{\cite{Rana2005ApJ...625..351R}}$ ($=1.7(3)\times10^{-10}~M_{\odot}yr^{-1}$, corrected to the Gaia EDR3 distance) is equal to  our value, within the error bars, they assumed $\dot{m}=1~g~cm^{-2}~s^{-1}$ that gave $f$ and $H_{sh}$ an order of magnitude larger than the ones we  derived, while we note that the larger $\dot{m}$ we adopted is consistent with the blackbody model. Fitting the intermediate polar mass model $\text{\citep{Suleimanov2016A&A...591A..35S, Suleimanov2019MNRAS.482.3622S}}$ to the 20-78 keV spectrum, $\text{\cite{Shaw2020MNRAS.498.3457S}}$ derived a total accretion flux and luminosity that are rather lower than these we reported here. If we ignore the reflection component from the  fit $C_C\#2$, the total flux is then estimated to be $7.5 \times 10^{-11}~ \text{erg} ~\text{s}^{-1} ~\text{cm}^{-2}$ that is consistent with the one ($8.5(\pm1) \times 10^{-11}~ \text{erg} ~\text{s}^{-1} ~\text{cm}^{-2}$) given by $\text{\cite{Shaw2020MNRAS.498.3457S}}$.

Another obvious difference between  previous results and ours is the estimation of the WD mass $M_{1}$, which maybe attribute to a different specific accretion rate adopted by them and the reflection effect involved in our model. If we ignore the reflection effect in spectral  fit $C_M\#2$, the \nustar\ + \xrt\ data give $M_{1} \sim$  1.0 $M_{\odot}$ that is closer to the estimation  of \cite{Rana2005ApJ...625..351R}, namely $M_{1}=1.2~M_{\odot}$ from multi-component plasma model and to the estimation of $\text{\cite{Ramsay2000MNRAS.314..403R}}$ of $M_{1}=0.98~M_{\odot}$ using the same method. Our $M_1$ determination is consistent with the lower limit of $0.82 M_{\odot}$ suggested by $\text{\cite{Ramsay2000MNRAS.314..403R}}$ from Fe K lines, also the radial velocity of narrow H$_{\alpha}$ lines favors this determination. The narrow emission lines arise from the heated hemisphere of the secondary \citep{Patterson1995PASP..107..307P} and yield a velocity amplitude of  $\lesssim$ 180 $km~s^{-1}$ \citep{Staubert1994A&A...288..513S, Watson1995MNRAS.273..681W, Patterson1995PASP..107..307P}. This velocity is  consistent with that of the inner Lagrangian $L1$ point ($\sim$ 180 $km~s^{-1}$) if the system has $M_1$= 0.85 $M_{\odot}$ and other parameters calculated in Section \ref{sub:param}, as the case in AM Her, where the narrow emission component origins from the $L1$ point \citep{Mukai1988MNRAS.232..175M}. If $M_1=1.2 (0.98)~M_{\odot}$, the minimal radial velocity $235 (200)~km~s^{-1}$ of the material in the secondary star is much larger than that observed. Fitting the intermediate polar mass model $\text{\citep{Suleimanov2016A&A...591A..35S,Suleimanov2019MNRAS.482.3622S}}$ to the 20-78
keV \nustar\ spectrum  \cite{Shaw2020MNRAS.498.3457S} derived a same WD mass $M_1 = 0.76_{-0.08}^{+0.09} M_{\odot}$ as ours within the errors, while from the observed egress duration, \cite{Mukai2003ApJ...597..479M} estimated for $M_1$ a low upper limit of $0.67~M_{\odot}$. We speculate that the radiation from the complex structure (e.g. the ring around the WD) maybe contaminate the flux and lower their estimation of $M_1$. In Table \ref{tab:mass_estimates}, we listed all the WD mass estimates for V1432 Aql in the literature.

\begin{table*}
\centering
\renewcommand\tabcolsep{3pt}
\caption{Mass Estimates for the White Dwarf in V1432 Aql} \label{tab:mass_estimates}
\begin{tabular}{cccc}
\hline\hline
M$_1$(M$_{\odot}$) & Data (Energy Band) & Method$^*$ & Reference\\
\hline
0.98  & \rxte~(2-6; 7.5-20~keV) & spectral fit  & \cite{Ramsay2000MNRAS.314..403R} \\
$<$0.67 & \rxte~(3-10~keV) & eclipsing analysis & \cite{Mukai2003ApJ...597..479M} \\
1.2(1) & \rxte~(2-10~keV) & spectral fit & \cite{Rana2005ApJ...625..351R} \\
0.76$_{-0.08}^{+0.09}$ & \nustar~(20-78~keV) & spectral fit & \cite{Shaw2020MNRAS.498.3457S} \\
 0.85(7) & \swift~+\nustar~(0.3-78~keV) & spectral fit & this work \\
\hline\hline
\end{tabular}
\begin{minipage}{\textwidth}
\vspace{5pt}
Notes: $^*$ refer to the corresponding reference for more details.\\
\end{minipage}
\end{table*}

\section{Discussion and Conclusions}\label{sec:concl}
~~~~From the analysis of the orbit-folded X-ray light curves, we found that the light curves present a profile with a low-intensity state for almost half an orbital period, a dip structure at the 0.6 phase, and a pulse at 0.75 phase, which indicate that only one accretion region was active during the observation and that the system has a complex environment. Although the profile is very similar to that obtained by \rosat~$\text{\cite{Rana2005ApJ...625..351R}}$, the   light curves have an orbital modulation, as the case in \sax, which is  different from the spin modulation in the \rosat\ and \xmm\ data, and we speculate that there are two different accretion geometries for V1432 Aql. In polars the accretion stream is channeled directly by the magnetic field lines at the threading point, this  stream-fed accretion geometry is found to be at work in other APs (i.e. BY Cam, CD Ind and J0838). While the inefficient and variable, depending on beat phase, threading in APs causes another accretion geometry, wherein the accretion stream can travel  some distance around the WD and the magnetic field  may channel the matter from a  ring of material, and this  is referred to as ring-fed accretion geometry, and is similar to that observed in the intermediate polars.  This disk-like structure manifested itself in the Doppler tomography and  caused the optical modulation of the mid-eclipse times and magnitudes.

The spectral analysis of the joint \nustar\ and \xrt\ data unambiguously present a reflection effect in V1432 Aql and a complex absorption environment, which explained the large Fe K$\alpha$ equivalent width EW $\sim 200$ eV, as has already been suggested before \text{\citep[e.g.][]{Rana2005ApJ...625..351R}}. Based on the wide-band X-ray data, we found that the emission energy maintains a balance between the soft and hard X-rays, which is consistent with the prediction by the standard model for polars, and that the system has a soft X-ray temperature $\sim 50$ eV observed commonly in polars.  Considering a exponential distribution of the heated area as a function of the temperature, we estimated the fractional accretion area as $\sim 0.034\%$ and the mean specific accretion rate as $\sim 3.8$ g cm$^{-2}$ s$^{-1}$, which predict a low shock height $\sim 0.02 R_{\odot}$ and justify the validity of the post-shock model where each bremsstrahlung component cools at a constant pressure and gravity. We also estimated the WD mass $M_1 \sim$  0.85 $M_{\odot}$ and found that this estimation is consistent with the radial velocity of narrow emission lines and the measurement of $\text{\cite{Shaw2020MNRAS.498.3457S}}$ based on the same \nustar\ data.

\normalem
\begin{acknowledgements}
We thank the anonymous referee for numerous useful comments that helped improve the manuscript. This work is supported by the National Natural Science Foundation of China (No. 11933008 and No. 11922306) and Chinese Academy of Sciences Interdisciplinary Innovation Team. This work has made use of data from the European Space Agency (ESA) mission \gaia\ (\url{https://www.cosmos.esa.int/gaia}), processed by the \gaia\ Data Processing and Analysis Consortium (DPAC, \url{https://www.cosmos.esa.int/web/gaia/dpac/consortium}). Funding for the DPAC
has been provided by national institutions, in particular the institutions participating in the \gaia\ Multilateral Agreement. The work is also based on observations obtained with the NuSTAR mission, a project led by the California Institute of Technology (Caltech), managed by the Jet Propulsion Laboratory and funded by NASA. We thank the \nustar\ Operations, Software and Calibration teams and the \swift\ Operations team for support with the execution and analysis of these observations. 
\end{acknowledgements}

\bibliographystyle{raa}
\bibliography{bibtex}

\end{document}